\documentclass[11pt]{article}
\usepackage{amsmath,amsthm,amssymb,euscript,epsf,epsfig}
\usepackage{array}
\usepackage{fancybox}
\usepackage[usenames]{color}
\usepackage{amsfonts,bm}
\usepackage{graphicx}
\usepackage{enumerate}

\usepackage[
      colorlinks=true,
      linkcolor=blue,
      urlcolor=blue,
      filecolor=blue,
      citecolor=blue,
      pdfstartview=FitH,
      pdftitle={The supergravity fields for a D-brane with a travelling wave from string amplitudes},
      pdfauthor={William Black, Rodolfo Russo, David Turton},
      pdfsubject={},
      pdfkeywords={Black holes, D-branes, Boundary state},
      pdfpagemode=UseNone,
      bookmarksopen=true
      ]{hyperref}
\usepackage[all]{hypcap}     

\numberwithin{equation}{section}

\newcommand{\be}{\begin{equation}}
\newcommand{\ee}{\end{equation}}
\newcommand{\bea}{\begin{eqnarray}\displaystyle}
\newcommand{\eea}{\end{eqnarray}}

\newcommand{\nn}{\nonumber}

\def\eq#1{(\ref{#1})}

\def\beq{\begin{equation}}
\def\eeq{\end{equation}}
\def\beqa{\begin{eqnarray}}
\def\eeqa{\end{eqnarray}}
\def\bet{\begin{tabular}}
\def\eet{\end{tabular}}

\def\a{\alpha}

\def\G{\Gamma}
\def\e{\epsilon}
\def\vep{\varepsilon}

\def\f{\phi}

\def\m{\mu}

\def\n{\nu}

\def\r{\rho}
\def\k{\kappa}

\def\s{\sigma}
\def\t{\tau}

\def\p{\psi}

\def\cA{{\cal A}}  
  
\def\cG{{\cal G}}  
  
\def\cM{{\cal M}} \def\cN{{\cal N}}

\def\mR{ \mathbb{R}} 
\def\mZ{ \mathbb{Z}}

\def\one{{\hbox{\kern+.5mm 1\kern-.8mm l}}}
\def\zero{{\hbox{0\kern-1.5mm 0}}}

\definecolor{orange}{rgb}{1,0.5,0}

\newcommand{\bra}[1]{{\langle {#1} |\,}}
\newcommand{\ket}[1]{{\,| {#1} \rangle}}
\newcommand{\braket}[2]{\ensuremath{\langle #1 | #2 \rangle}}

\def\d{ \partial } 
\def\zb{{\bar z}}

\newcommand{\dbyd}[1]{\ensuremath{ \frac{\d}{\d {#1}}}}

\newcommand{\T}[3]{\ensuremath{ #1{}^{#2}_{\phantom{#2} \! #3}}}		

\newcommand{\tr}{\operatorname{tr}}

\def\ha{\frac{1}{2}}
\def\tha{\tfrac{1}{2}}

\def\pt{\widetilde{\psi}}
\def\at{\widetilde{\a}}

\def\d{\partial}
\def\db{\bar\partial}
\def\delb{\bar\partial}

\def\delb{\bar\partial}
\newcommand{\ex}[1]{{\rm e}^{#1}} 
\def\ii{{i}}

\newcommand{\ap}{\ensuremath{\alpha'}}

\newcommand{\bean}{\begin{eqnarray*}}
\newcommand{\eean}{\end{eqnarray*}}

\newcommand{\hsp}{\hspace{0.5cm}}

\begin{document}

\hfill \hbox{QMUL-PH-10-07}

\vskip 1.5cm

\centerline{
{\LARGE \bf  The supergravity fields for a D-brane with a travelling wave }}
\vskip 0.5cm
\centerline{
{\LARGE \bf  from string amplitudes}}

\vskip 1.8cm

\centerline{    
  {\large {\bf William Black\footnote{w.black@qmul.ac.uk}, Rodolfo Russo\footnote{r.russo@qmul.ac.uk}, David Turton\footnote{d.j.turton@qmul.ac.uk} }  }  }

\vskip 1.2cm

\begin{center}
{Queen Mary University of London\\
Centre for Research in String Theory \\ 
Department of Physics\\
Mile End Road\\
London E1 4NS, UK\\
}
\end{center}

\vskip 1.8cm

\centerline{\bf Abstract} \vskip 0.5cm

We calculate the supergravity fields sourced by a D-brane with a null
travelling wave from disk amplitudes in type IIB string theory
compactified on $T^4 \times S^1$. The amplitudes reproduce all the
non-trivial features of the previously known two-charge supergravity
solutions in the D\nobreakdash-\hspace{0pt}brane/momentum
duality frame, providing a direct link
between the microscopic bound states and their macroscopic
descriptions.

\vskip 2.5cm


\thispagestyle{empty}

\vfill
\eject

\section{Introduction and Discussion}
\setcounter{footnote}{0}
D-branes in string theory appear both as classical solutions of the
supergravity low-energy effective action and as fundamental objects on
which open strings can end. It is well known that mixed open/closed
string amplitudes can be used to derive information about the
classical solution from the microscopic
description~\cite{garousi96,hashimoto96}. In particular, the one-point
functions of closed string states from the disk provide a direct way
to compute, in the small coupling regime, the backreaction of the
D-branes considered~\cite{DiVecchia:1997pr,Billo:1998vr,DiVecchia:1999uf,DiVecchia:1999rh}.

Recently it has been shown for two-charge D1-D5 configurations that
the same direct link between the microscopic and the macroscopic
descriptions holds~\cite{Giusto:2009qq}.  Two-charge D-brane
configurations have attracted much attention as they represent a
simple and tractable system within which (small) black holes in
string theory may be studied~\cite{Sen:1995in,Das:1996ug,Lunin:2001fv,Lunin:2001jy,Dabholkar:2004yr,Mathur:2005zp,Skenderis:2008qn,Sen:2009bm,Chowdhury:2010ct}. A
natural question is whether these studies can be extended to
three-charge D1-D5-P configurations~\cite{Strominger:1996sh,Bena:2007kg}, since the supergravity
description of these configurations is that of a macroscopic black
hole.  As a further step towards this goal,
here we study two-charge configurations involving a momentum
charge.

In this letter we calculate the supergravity fields sourced by a D-brane
carrying momentum charge in the form of a null right (or left)
moving wave, and show that
the fields sourced by this bound state reproduce 
the non-trivial features of the supergravity solutions which are U-dual to the 
fundamental string solution of~\cite{Dabholkar:1995nc,Callan:1995hn}.
In particular we describe in detail the calculation in the D5-P duality frame.
The world-sheet calculation employs the fact that these D-brane configurations admit 
an exact CFT description~\cite{Bachas:2002jg} in which the travelling wave on the
D-brane can be included in the world-sheet action for the open strings
in a tractable way. We use the boundary state describing a D-brane
with a travelling wave~\cite{Hikida:2003bq,Blum:2003if,Bachas:2003sj}
to compute the disk one-point functions for emission of massless closed string
states, and we read off the various supergravity fields. Contrary to what happens in the D1-D5
frame~\cite{Giusto:2009qq}, the string computation in these duality
frames yields the full integrals over the D-brane profile appearing in the classical solutions. 
This is possible because the profile function parametrizing the solutions
arises as a condensate of massless open strings related to the
physical shape of the D-brane, which can be included exactly in the
string world-sheet action.

The direct link between microscopic D-brane configurations and supergravity
solutions might also shed further light on the entropy of two charge
systems in string theory. 
It was recently proposed~\cite{Sen:2009bm}
that the macroscopic entropy of a two-charge configuration should
be defined to be the sum of the contributions of small black hole solutions and
horizonless smooth classical solutions. In this language the term `smooth classical solutions' does not 
include solutions which are singular due to delta-function sources,
and the scaling arguments of~\cite{Sen:2009bm} applied to the D\nobreakdash-\hspace{0pt}brane/momentum
duality frame show that
$\ap$-corrections to the supergravity action cannot produce small black holes with a non-zero horizon area.

Here we observe that the supergravity solutions which are sourced by the 
microscopic D-brane bound states are necessarily singular at the two-derivative level:
the one-point functions on the disk discussed in this letter provide
the asymptotic behaviour of the solutions, and the nonlinear part of
the standard supergravity equations of motion determines the
background in the interior, leading to the singular backgrounds
obtained by dualising the fundamental string solution. Of course, it
might still be possible to recover a fully smooth field configuration
starting from the same data provided by the disk one-point functions
if one includes $\ap$-corrections to the supergravity equations of
motion.

This letter is organised as follows. In Section~2 we review the
two-charge supergravity solutions in the D1-P and D5-P duality frames
and present the terms in the perturbative expansion that we
reproduce from the string amplitudes. In Section~3 we derive the
boundary state for these D-brane configurations, compute the one-point
functions of the massless closed string states and read off the
gravitational backreaction, at weak coupling, of the D-brane
configuration.

\section{Two-charge system in the D1-P and D5-P duality frames}

We work in type IIB string theory on $\mathbb{R}^{4,1}\times S^1\times
T^4$ using the light-cone coordinates $u = \left( t+y \right) ,
~ v = \left( t - y \right) \, $ constructed from the time and $S^1$ directions.  The indices $(I,J,\ldots)$ refer
collectively to the other eight directions which we then split into
the $\mathbb{R}^{4}$ directions labelled by $(i,j,\ldots)$ and those along the
$T^4$ labelled by $(a,b,\ldots) \,$.

The family of classical supergravity solutions in which we are interested
describe two-charge D-brane bound states~\cite{Lunin:2001fv,Lunin:2001jy,Lunin:2002bj,Lunin:2002iz,Kanitscheider:2007wq}
and are connected through T and S dualities to the
solution describing a multi-wound
fundamental string with a purely right (or left) moving wave~\cite{Dabholkar:1995nc,Callan:1995hn}, 
smeared along the $T^4$ directions and along $y$~\cite{Lunin:2001fv,Lunin:2001jy}. In the
D1-P duality frame, we take the D1-brane to be wrapped $n_w$ times around~$y$; letting the length of the $y$ direction be $2 \pi R$, the brane then has overall extent $L_T = 2 \pi n_w R$ and we use $\hat{v}$ for the corresponding world-volume coordinate on the D-brane, having periodicity $L_T$. The non-trivial fields are the metric, the dilaton
and the R-R 2-form gauge potential:
\bea\label{D1Psol}
ds^2 &=&   H^{-\frac{1}{2}} dv \;\Big(\!\! - du +  K  dv +
2 A_{I}  dx^I \Big) +  H^{\frac{1}{2}} dx^I dx^I~,  \\ 
e^{2 \Phi} &=&  g_s^2 H~, \hsp
C^{(2)}_{uv} ~=~ - \tha (H^{-1} -1)~, \hsp
C^{(2)}_{vI} ~=~ - H^{-1} A_I ~, \nn
\eea
where the harmonic functions take the form 
\beq \label{eq:F1P-HAK}
H = 1 + \frac{Q_1}{L_T} \int\limits_0^{\,\,L_T}
\frac{d\hat{v}}{|x_i-f_i(\hat{v})|^2} ~, \hsp 
A_I = -\frac{Q_1}{L_T} \int\limits_0^{\,\,L_T} \frac{d\hat{v}
  \dot{f}_I(\hat{v})}{|x_i-f_i(\hat{v})|^2} ~, \hsp
K = \frac{Q_1}{L_T} \int\limits_0^{\,\,L_T} \frac{d\hat{v} |\dot{f}_I (\hat{v})|^2}
{|x_i-f_i(\hat{v})|^2} ~,
\eeq
where $f_i(\hat{v}+L_T)=f_i(\hat{v})$ and where $\dot f$ denotes the derivative of $f$ with respect to
$\hat{v}$. The functions $f_I$ describe classically the null travelling wave
on the D-string. $Q_1$ is proportional to $g_s$ and to the D-brane winding number $n_w$
and is given by
\be 
Q_1 ~=~ \frac{(2\pi)^4 n_w g_s (\ap)^3}{V_4}~.
\label{eq:QD1}
\ee
T-dualising to the D5-P duality frame and using the symmetry of the IIB equations of motion to reverse the sign of $B$ and $C^{(4)}$, we obtain the fields
\bea
ds^2  &=&  H^{-\frac{1}{2}} dv \;\Big(\!\! - du + \left( K- H^{-1}
|A_a|^2 \right) dv + 2 A_{i} dx_i\Big) + H^{\frac{1}{2}} dx^i dx^i +
H^{-\frac{1}{2}} dx^a dx^a\;, \cr
e^{2 \Phi} &=&  (g_s')^2 H^{-1}~, \qquad ~~ B_{va} ~=~ - H^{-1} A_a~,
\label{D5Psol} \\ 
C^{(4)}_{vbcd} &=&       - H^{-1} A_a\e_{abcd}~, \quad
C^{(6)}_{vi5678} ~=~   - H^{-1} A_i~, \quad
C^{(6)}_{uv5678} ~=~   - \tha \left( H^{-1}- 1 \right), \nn
\eea
where $g_s'$ is the string coupling in the new duality frame and $\e_{abcd}$ is the alternating symbol with $\e_{5678}=1$.
The effect of rewriting the functions in~\eqref{eq:F1P-HAK} in terms
of D5-P frame quantities is to substitute the D1 with the D5 charge,
$\,Q_1\to Q_5 = g_s' n_w \ap\,$. From now on, we drop the prime and refer to the 
D5-P frame string coupling as $g_s$.

From the large distance behaviour of the $g_{vv}$ component of
the metrics above, one can read off how the momentum charge is related to the
D-brane profile function $f$. For instance, in the D1-P
frame we have
\be\label{nnp}
\frac{n_w}{L_T} \int\limits_0^{\,\,L_T} |\dot{f}|^2 d\hat{v} = \frac{g_s n_p
  \ap}{R^2}~, 
\ee
where $n_p$ is the Kaluza-Klein integer specifying the momentum along
the compact $y$ direction. 
From a statistical point of view~\cite{Mathur:2005zp}, the
typical two-charge bound state with fixed D1 and momentum charges has
a profile $f$ consisting of Fourier modes of average frequency $\sqrt{n_w
n_p}$. Then~\eqref{nnp} implies that the typical profile wave has an amplitude of order
$\sqrt{g_s}$. Despite this potential $g_s$ dependence, we always keep track of $f$ exactly and
expand in the D-brane charges $Q_i$. From the point of view of the string
amplitudes, this means that we are resumming all diagrams with
open string insertions describing the D-brane profile, but that we are considering
only the disk level contribution.

From now on, for concreteness we present the calculation in the D5-P frame 
and we focus on the field components that vanish in the absence
of a wave; the calculations of the remaining components are analogous. 
We canonically normalize the metric, B-field and R-R fields: 
\be
g=\eta+2 \kappa \;\! \hat{h} \, , \qquad B=\sqrt{2} \k \;\! \hat{b} \,, \qquad C = \sqrt{2} \kappa \;\! \hat{C}
\ee
where as usual, $\kappa=2^3 \pi^{7/2} g_s(\ap)^2$. We then expand the 
relevant components of~\eqref{D5Psol} for small $Q_5$, keeping only linear order terms, which yields the field 
components that we shall reproduce from the disk amplitudes:
\be 
\hat{h}_{vi}
~=~  \frac{Q_5}{2 \k L_T}  
\int\limits_0^{\,\,L_T} \frac{- \dot{f}_i \,d\hat{v}}{|x_i-f_i(\hat{v})|^2}\;,~~ 
\hat{h}_{vv} 
~=~ \frac{Q_5}{2 \k L_T}  
\int\limits_0^{\,\,L_T} \frac{|\dot{f}|^2 \, d\hat{v} }{|x_i-f_i(\hat{v})|^2} \;, ~~
\hat{b}_{va} 
~=~ \frac{Q_5}{\sqrt{2} \k L_T} 
\int\limits_0^{\,\,L_T}
\frac{ \dot{f}_a \, d\hat{v}}{|x_i-f_i(\hat{v})|^2} \,, \nn
\ee 
\be \label{eq:sugralinear} 
\hat{C}^{(4)}_{vbcd}  
~=~ \frac{Q_5}{\sqrt{2} \k L_T}  
\int\limits_0^{\,\,L_T}
 \!d\hat{v} \,
\frac{ \dot{f}_a \e_{abcd} }{|x_i-f_i(\hat{v})|^2}~,\qquad
\hat{C}^{(6)}_{vi5678}  
~=~ \frac{Q_5}{\sqrt{2} \k L_T} 
\int\limits_0^{\,\,L_T}
 \!d\hat{v}
\,\frac{\dot{f}_i}{|x_i-f_i(\hat{v})|^2}~.
\ee
Similar expressions are easily
derived in the D1-P frame from~\eqref{D1Psol}.

\section{Classical fields from string amplitudes}

\subsection{World-sheet boundary conditions}

The key ingredients of our string computation are the boundary conditions which must be imposed upon the world-sheet fields
of a string ending on a D-brane with a travelling wave, which we now review. We consider a Euclidean world-sheet
with complex coordinate $z = \exp (\t-i\s) $ such that $\t \in \mR$
and $\s \in [0,\pi] \,$. We first review the boundary conditions applicable for a D-brane wrapped only once around $y$ and later account for higher wrapping numbers.

We begin with the following world-sheet action for the superstring coupled to a background gauge field $A^{\m}$ on a D9-brane following 
\cite{HaggiMani:2000uc,Blum:2003if}:
\beq 	
	\label{eq:open_string_action}
	S~=~S_0+S_1~,
\eeq
where $S_0$ and $S_1$ are the world-sheet bulk and boundary actions respectively,
\begin{subequations}
\begin{align}
	S_0~=~&\, \frac{1}{2\pi \ap} \int\limits_M d^2z\, \bigg( \d X^{\m}\delb X_{\m}+\p^{\m} \delb \p_{\m}+\pt^{\m} \d \pt_{\m}\bigg) \, ,\\
	S_1~=~&i \int\limits_{\d M} dz \,\bigg( A_{\m}(X) \left( \d X^{\m} + \bar{\d} X^{\m}\right) 
	-\frac{1}{2} \big( \p^{\m}+\pt^{\m}\big) F_{\mu\nu} \big(\p^{\n}-\pt^{\n}\big) \bigg)
\end{align}
\end{subequations}
and $F_{\mu\nu} = \d_{\m}A_{\n}-\d_{\n}A_{\m}$ is the abelian field strength. In the absence of a boundary the action $S_0$ would be invariant under the 
supersymmetry transformations
\begin{equation}\label{eq:susy_variations}
\delta X^{\m} =\vep \p^{\m}+\tilde{\vep}\pt^{\m} ~ ,~~~ 
\delta \p^{\m}=-\vep \d X^{\m} ~,~~~ 
\delta \pt^{\m}=-\tilde{\vep} \delb X^{\m} ~
\end{equation}
however the presence of the boundary breaks the $\cN=2$
world-sheet supersymmetry to $\cN=1$ supersymmetry.  When we include $S_1$, the total action $S_0 + S_1$ preserves $\cN=1$ supersymmetry only up to the boundary conditions~\cite{HaggiMani:2000uc}, which we impose at $z=\bar{z}$. Defining
\be
E_{\m\n} ~=~ \eta_{\m\n}+ 2 \pi \ap F_{\m\n}~, 
\ee
varying the above action yields the boundary
conditions~\cite{HaggiMani:2000uc}
\begin{subequations}
\begin{gather}
	\left[ E_{\m\n}\pt^{\n} ~=~ \eta E_{\n\m} \psi^{\n} \right]_{z = \zb} ~,  \label{eq:bcbosons} \\
	\left[ E_{\m\n}\db X^{\n} - E_{\n\m} \d X^{\n}
 -  \eta E_{\n\r,\m} \pt^{\n} \psi^{\r} - E_{\m\n,\r} \psi^{\n}\psi^{\r} + E_{\n\m,\r} \pt^{\n} \pt^{\r}\right]_{z = \zb} ~=~ 0 \, ,
 \label{eq:bcfermions}
\end{gather}
\end{subequations}
where $\eta$ takes the value $1$ or $-1$ corresponding to the NS and R
sectors respectively. By applying the supersymmetry transformations
(\ref{eq:susy_variations}) to the action (\ref{eq:open_string_action})
and employing these boundary conditions, one finds that
(\ref{eq:open_string_action}) is invariant under the $\cN=1$ supersymmetry
generated by these transformations
with the constraint $\vep=\eta \tilde{\vep}$.

For the systems under consideration the gauge field takes a plane-wave
profile and so $A^{\m}$ will be a function only of the bosonic field
${V}=(X^0-X^9)$, where $X^0$ is the string coordinate along time
and $X^9$ indicates the compact $y$ direction. 
A physical gauge field can be written as $A^{I}({V})$, where we set to
zero the light-cone components. Then the non-vanishing components of
$E_{\m\n}$ take the form
\be
E_{uv} = E_{vu} = - \ha \, , \qquad  E_{IJ} = \delta_{IJ} \, , \qquad
E_{Iv} = - E_{vI}= \dot{f}_I({V})~, 
\ee
where we have defined $f_{I} = - 2 \pi \ap A_{I}$. 

We can rewrite the fields appearing~\eq{eq:bcbosons}
and~\eq{eq:bcfermions} in modes by using the expansions
\bea \label{eq:modes}
&& X^{\m}(z,\zb)  ~=~ x^{\m} 
-i \sqrt{\frac{\ap}{2}} \alpha_0^\m \ln z 
-i \sqrt{\frac{\ap}{2}} \at^{\m}_0 \ln \bar{z}
+ i \sqrt{\frac{\ap}{2}} \sum_{m\ne 0} \frac{1}{m}
\left( \frac{\a^{\m}_m}{z^m} + \frac{\at^{\m}_m}{\zb^m} \right) , \\
&& \psi^{\m}(z) ~=~ \sqrt{\frac{\ap}{2}} 
\sum_{r\in \mZ +\n} \frac{\p_r^{\m}}{z^{r+\ha}}~, \qquad 
\pt^{\m}(\zb) ~=~ \sqrt{\frac{\ap}{2}} 
\sum_{r\in \mZ +\n} \frac{\pt_r^{\m}}{\zb^{r+\ha}} ~,
\label{eq:psi_modes}
\eea
where $\n = 0$ and $\tha$ for R and NS respectively. Note however that
in our case the presence of a non-constant field strength $F_{\m\n}$ makes
the boundary conditions nonlinear in the oscillators. We will see
that, for the amplitudes in which we are interested, only the linear terms
contribute.

As usual, we can change from the open string picture to the closed
string picture, and derive the boundary
conditions describing a closed string emitted or absorbed by the
D-brane.  This has the effect of
\be \label{eq:openclosed}
	\a_n^{\m} \to -\a_{-n}^{\m}~ , \qquad \qquad  \p_r^{\m} \to i\p_{-r}^{\m}~~  \qquad \qquad  \forall\, \m,n,r~.
\ee
We can then obtain the boundary conditions for a lower dimensional
D-brane by performing a series of T-dualities; after these
transformations, the components of $f$ along the dualised coordinates
describe the profile of the brane. We perform four or eight
T-dualities in order to obtain the boundary conditions appropriate for
a D5 or a D1-brane, for instance in order to move from the D9 frame
to the D5-P frame we T-dualise along each $x^i$ which sends
\be \label{eq:Txi}
\at_n^{i} \to - \at_{n}^{i}~ , \qquad \qquad \pt_r^{i} \to - \pt_{r}^{i}~ .
\ee
By following the procedure outlined above, we can summarise the
boundary conditions for the closed string oscillators as follows
\be \label{eq:D5-Poscbcs} 
\pt^{\m}_r  ~=~   i \eta \, \T{R}{\m}{\n} \psi^{\n}_{-r} ~ +
\ldots  ~, \qquad \qquad
\at^{\m}_n ~=~  - \T{R}{\m}{\n} \a^{\n}_{-n} ~ + \ldots~,
\ee
where `$\ldots$' indicates that we ignore terms which are higher than linear order in
the oscillator modes. We shall justify this below \eq{eq:NSNScoupling}. The reflection matrix $R$ is obtained
from~\eq{eq:bcbosons} and \eq{eq:bcfermions} by performing the
transformations \eq{eq:openclosed} and \eq{eq:Txi} and replacing $V$ by its
zero-mode $v$:
\be
\T{R}{\m}{\n}({v}) = \T{T}{\m}{\r} (E^{-1})^{\r\s} E_{\n\s}~,
\ee
where the matrix $T$ performs the T-duality \eq{eq:Txi},
i.e.~it is diagonal with values $-1$ in the $x^i$ directions and $1$
otherwise. $R$ has the lowered-index form
\be \label{eq:RD5p}
R_{\m\n}({v}) ~=~  \eta_{\m\r}\T{R}{\r}{\n}({v}) ~=~ \left( \begin{array}{cccc} 
   	- 2 |\dot{f}({v})|^2   &     - \ha    &   2\dot{f}^i({v})   &  2\dot{f}^a({v})  \\
          	- \ha                &      0      &          0              &         0         \\
 	  2 \dot{f}^i({v})     &      0      &    \!\! - \one         &         0         \\
  	 - 2 \dot{f}^a({v})     &      0      &          0             &   \!  \one
\end{array} \right) ~.
\ee

We refer the reader to~\cite{Hikida:2003bq,Blum:2003if,Bachas:2003sj} for a
detailed discussion of the boundary state describing a D-brane with a
travelling wave. For our purposes it is sufficient to know the
linearized boundary conditions for the non-zero modes~\eqref{eq:D5-Poscbcs} 
that the boundary state must satisfy, and to construct explicitly only the 
zero-mode structure of the boundary state. Addressing firstly the bosonic sector, 
the boundary conditions on the zero modes are
\be \label{eq:D5-Pbozonzeromodes}
 {p}_v  + \dot{f}^i({v}) \, {p}_i   ~=~ 0  \, ,  \qquad \quad
 {p}_u    ~=~ 0 \, , \qquad \quad
 {p}_a    ~=~ 0 \, , \qquad \quad
 {x}^{i} ~=~ f^i({v})  ~
\ee
where the first three equations follow directly from~\eqref{eq:D5-Poscbcs} and the fourth equation must be included 
to account for the T-duality transformations.
The first equation in \eq{eq:D5-Pbozonzeromodes} may be represented as
$~i \dbyd{v} = \dot{f}^i(v) {p}_i~$ and similarly the last constraint may be represented as $~i \frac{\partial}{\partial p_i} = f^i(v)\,$. Then the
boundary state zero-mode structure in the $t$, $y$ and $x^i$ direction
is
\beq
\int dv\, du \int \frac{d^4p_i}{(2 \pi)^4} e^{-i p_i
f^i({v}) } \ket{p_i} \ket{u} \ket{v}~.
\eeq

So far we have essentially discussed a D-brane with a travelling wave in a noncompact
space; we next generalise this description to the case of compact $y$ and higher
wrapping number.  One may view a D-brane wrapped $n_w$ times along the
$y$-direction as a collection of $n_w$ different D-brane strands with
a non-trivial holonomy gluing these strands together. This approach
was developed in~\cite{Duo:2007he,DiVecchia:2007dh} for the case of
branes with a constant magnetic field. 

In the presence of a null
travelling wave with arbitrary profile $f(V)$, the individual boundary
states of each strand will differ in their oscillator part and not
just in their zero-mode part described above.  However, we are
interested in the emission of massless closed string states, which have zero momentum
and winding along all compact directions. 
In this sector the full
boundary state is simply the sum of the boundary states for each
constituent, along with the condition that the value of the function
$f$ at the end of one strand must equal the value of $f$ at the
beginning of the following strand.
We label the 
strands of the wrapped D-brane with the integer $s$;
then restricting to the sector of closed strings with trivial winding ($m$)
and Kaluza-Klein momentum ($k$), the boundary state takes the
following form:
\be \label{eq:BdyStateStrands}
\ket{D5;P}^{k,m=0} =  - \frac{ \k \, \t_{5} }{2} 
\sum_{s=1}^{n_w} \int du \int\limits_0^{2 \pi R} d{v} \int \frac{d^4p_i}{(2 \pi)^4} 
e^{-i p_i f^i_{(s)}({v}) } 
\ket{p_i} \ket{u} \ket{v} 
\ket{D5;f_{(s)}}_{X,\psi}^{k,m=0}~,
\ee
where $\t_{5}= [(2\pi \sqrt{\ap})^5 \sqrt{\ap}g_s]^{-1}$ is the physical tension of a D5-brane. 
We have written explicitly only the bosonic zero-modes along $t$,
$y$ and the $x^i$ directions and we denote by $\ket{D5;f_{(s)}}_{X,\psi}^{k,m=0}~$
the remaining part of the boundary state. 
The range of integration over $v=t-y$ follows from the periodicity condition of the
space-time coordinate $y$.

We next address the fermion zero modes in the R-R sector.
Letting $A,B,...$ be 32-dimensional indices for spinors in ten
dimensions\footnote{For the spinors and the charge conjugation matrix, 
we use the conventions of \cite{DiVecchia:1997pr}.}, and letting $|A\rangle
|{\widetilde B}\rangle$ denote the ground state for the Ramond fields
$\psi^\mu(z)$ and ${\widetilde \psi}^\mu({\bar z})$ respectively,
the R-R zero mode boundary state in the $(-\tfrac{1}{2},-\tfrac{3}{2})$
picture (before the GSO projection) takes the form
\beq
|D5;P\rangle _{\psi,0}^{(\eta)} = {\cal M}_{AB}^{(\eta)}\,\ket{A}_{-\frac{1}{2}} \ket{\widetilde B}_{-\frac{3}{2}}
\label{eq:RRB0}
\eeq
where ${\cal M}$ satisfies the following equation \cite{DiVecchia:1997pr},
\be \label{eq:	}
\Gamma_{11} \, \cM \, \Gamma^\mu - i \eta\,\T{R}{\m}{\n}\,\left( \Gamma^{\nu} \right)^T \cM ~=~ 0 \, .
\ee
A solution to this equation for the case of our reflection matrix $R$ \eq{eq:RD5p} is given by\footnote{The overall phase of $M$ is a matter of convention; see also \cite{DiVecchia:1999uf}.} 
\newline
\be
M = i \, C \left( \ha \G^{vu} +  \dot{f}^I(v)\G^{Iv} \right) \G^{5678}  \left( \frac{\one - i \eta \G_{11}}{1- i \eta} \right) . 
\ee
where $C$ is the charge conjugation matrix. The GSO projection has the effect of 
\be
|D5;P\rangle _{\psi,0} = \ha \left( |D5;P\rangle _{\psi,0}^{(1)} + |D5;P\rangle _{\psi,0}^{(-1)} \right) 
\ee
and so the zero mode part of the D5-P R-R boundary state for the strand with profile $f_{(s)}$ is
\bea \label{eq:RRB}
|D5;f_{(s)}\rangle _{\psi,0} = i \left[ C \left( \ha \G^{vu} +  \dot{f}_{(s)}^I(v)\G^{Iv} \right) \G^{5678}
\frac{1+\G_{11} }{2}\right]
\ket{A}_{-\frac{1}{2}} \ket{\widetilde B}_{-\frac{3}{2}} ~
\eea
which we can insert into the relevant part of the boundary state \eq{eq:BdyStateStrands}.

\subsection{Disk amplitudes for the classical fields}

We now calculate the fields sourced by the D5-P bound state by computing the disk one-point functions
for emission of a massless state, starting with the NS-NS fields. Since the states are massless they have 
non-zero momentum only in the four noncompact directions of the $\mathbb{R}^4$, 
i.e. they have spacelike momentum (see also~\cite{DiVecchia:1997pr}). The NS-NS one-point function thus takes the form
\be
\cA_{\rm NS}^{(\eta)}(k)  ~\equiv~ 
\bra{p_i=k_i}\bra{p_{v}=0}\bra{p_{u}=0}\bra{n_a=0}\cG_{\m\n}
\psi^{\mu}_{\frac{1}{2}}\, \widetilde{\psi}^{\nu}_{\frac{1}{2}} 
\ket{D5;P}^{k,m=0}~
\ee
where for an $S^1$ direction with radius $R$ we normalize the momentum
eigenstates as $\braket{n}{m}=2 \pi R \, \delta_{nm}$ and the position eigenstates as $\braket{x}{y} =
\delta(x-y)$. In terms of canonically normalized fields, $\cG_{\m\n}$ is given by
\be
\cG_{\m\n} = \hat{h}_{\m\n} + \frac{1}{\sqrt{2}} \hat{b}_{\m\n} + \frac{\f}{2\sqrt{2}} \left( \eta_{\m\n} - k_{\m}l_{\n} - k_{\n}l_{\m} \right) \,,
\ee
where $k_{\m}$ and $l_{\n}$ are mutually orthogonal null vectors.  The
contribution to the zero mode part of the amplitude from a single
strand with profile $f_{(s)}(v)$ is
\be \label{intre}
V_4 V_u \frac{ \k \, \t_{5} }{2} \int\limits_0^{2 \pi R} \! dv  \,  e^{- i k_i f_{(s)}^i (v) } ~,
\ee
where $V_u$ represents the infinite volume of the D-brane in the $u$
direction. Since we have used a delocalised probe ($p_v=0$), the string
amplitude contains an integral over the length of the strand of the
D-brane. In the classical limit $n_w$ is very large, the typical wavelength of the profile is much bigger than
$R$, and so $f$ is almost constant over each strand~\cite{Lunin:2001fv,Lunin:2001jy}. 
The contribution to the value of each supergravity field is thus~\eqref{intre} divided by the volume of the strand:
\be \label{eq:A_0^s}
\cA_0^{(s)}(k) ~=~ \frac{ \k \, \t_{5} }{2} \frac{1}{2 \pi R} \int\limits_0^{2 \pi R} \! dv  \,  e^{- i k_i f_{(s)}^i (v) }~.
\ee
The contribution from the $n_w$ different strands of the brane is therefore
\be
\cA_0(k) ~=~ \frac{ \k \, \t_{5} }{2} \frac{1}{2 \pi R} \sum_{s=1}^{n_w} \int\limits_0^{2 \pi R} \! dv  \,  e^{- i k_i f_{(s)}^i (v) }~,
\ee
and we combine the integrals over each strand to give the integral over the full
world-volume coordinate $\hat{v}$, giving
\be \label{eq:A_0}
\cA_0(k) ~=~ \frac{ \k \, \t_{5} }{2}  \frac{n_w}{L_T}
\int\limits_0^{\,\,L_T}
\! d\hat{v}  \,  e^{- i k_i f^i (\hat{v}) }~.
\ee
Adding in the non-zero modes, the coupling of the boundary state to the
NS-NS fields is
\be \label{eq:NSNScoupling}
\cA_{\rm NS}^{(\eta)}(k)  
 ~=~ 
 - i \eta \frac{\k \, \t_{5} \, n_w}{2 L_T} 
 \int\limits_0^{\,\,L_T}
 \! d\hat{v}  \,  e^{- i k_i f^i (\hat{v}) }   \cG_{\m\n} R^{\n\m}(\hat{v})~
\ee
where $R(\hat{v})$ is the obvious strand-by-strand extension of the
reflection matrix \eq{eq:RD5p}. 

We can now observe why we were justified in ignoring terms higher than linear order in the oscillator boundary conditions \eq{eq:D5-Poscbcs}. To arrive at the above result we substitute $\widetilde{\psi}^{\nu}_{\frac{1}{2}}$ for an expression involving only creation modes using \eq{eq:D5-Poscbcs}, and only the linear term can contract with the remaining annihilation mode to give a non-zero result.
A similar argument holds for the R-R amplitude.

The GSO projection has the effect of
\be
\cA_{\rm NS}(k) ~=~ \ha \left( \cA_{\rm NS}^{(1)}(k) - \cA_{\rm NS}^{(-1)}(k) \right) 
\ee
and we read off the canonically normalized fields of interest via 
\bea \label{vans}
\hat{h}_{v i}(k) &=&  \frac 12  \frac{\delta {\cal A}_{\rm NS}}{\delta \hat{h}^{v i} }~,  \qquad \quad
\hat{h}_{v v}(k) ~=~ \frac{\delta {\cal A}_{\rm NS}}{\delta \hat{h}^{v v}} \, ,  \qquad \quad
\hat{b}_{v a}(k) ~=~ \frac{\delta {\cal A}_{\rm NS}}{\delta \hat{b}^{v a}} ~.
\eea
The space-time configuration associated
with a closed string emission amplitude is obtained by multiplying the
derivative of the amplitude with respect to the closed string field by
a free propagator and taking the Fourier transform~\cite{DiVecchia:1997pr}. 
In general for a field $a_{\mu_1\ldots\mu_n}$ we have
\beq\label{fourier}
a_{\mu_1\ldots\mu_n}(x) = \int \frac{d^4 k}{(2\pi)^4}
\left(-\frac{\ii}{k^2}\right)  
a_{\mu_1\ldots\mu_n}(k)\,
\ex{\ii k x} ~,
\eeq
with $a_{\mu_1\ldots\mu_n}(k)$ given in terms of derivatives of ${\cal
A}$ as in~(\ref{vans}). Using the identity
\begin{equation}
\int \frac{d^4 k}{(2\pi)^4}  \frac{e^{i k^i (x^i - f^i) }}{k^2}  ~=~  \frac{1}{4 \pi^2} \frac{1}{|x^i - f^i|^2}\, 
\end{equation}
and the relation
\be
Q_5 ~=~ \frac{2 \k^2 \, \t_{5} \, n_w}{4 \pi^2}~,
\ee
we obtain
\bea
\hat{h}_{v i} &=&  \frac{Q_5}{2\k L_T}
\int\limits_0^{\,\,L_T}
\! \frac{ - \dot{f}_i \, d\hat{v} }{|x^i - f^i|^2} , \quad
\hat{h}_{v v} ~=~ \frac{Q_5}{2\k L_T} 
\int\limits_0^{\,\,L_T}
 \! \frac{|\dot f|^2 \, d\hat{v} }{|x^i - f^i|^2} ,\quad
\hat{b}_{v a} ~=~  \frac{Q_5}{\sqrt{2}\k L_T} 
\int\limits_0^{\,\,L_T}
 \! \frac{\dot{f}_a \, d\hat{v} }{|x^i - f^i|^2} \nn 
\eea
in agreement with \eq{eq:sugralinear}.

We next calculate the coupling between the R-R zero mode boundary state and the on-shell R-R potential state \cite{DiVecchia:1999uf,DiVecchia:1997pr,Billo:1998vr}:
\begin{eqnarray}
\bra{\hat{C}_{(n)}} &=& 
{}_{-\frac{1}{2}}\bra{\widetilde{B}, \tfrac{k}{2} } ~{}_{-\frac{3}{2}}\bra{A, \tfrac{k}{2}} 
\left[ C\Gamma^{\mu_1\ldots\mu_n}\frac{\one - \G_{11}}{2} \right]_{AB} \,
\frac{(-1)^n}{4\sqrt{2}\,n!}~ \hat{C}_{\mu_1\ldots\mu_n} ~
\label{eq:RRonshell}
\end{eqnarray}
where the numerical factor contains an extra factor of $\tha$ to account for the fact that we are not using the full superghost expression.
Using the fact (see e.g.~\cite{Billo:1998vr}) that
\begin{equation}
\label{bs47}
\left (\bra{A}\bra{\tilde B}\right)\,
\left(\ket{D}\ket{\tilde E}\right) =
  - \braket{A}{D}~\braket{\tilde B}{\tilde E}
  = - (C^{-1})^{AD}(C^{-1})^{BE} \, ,
\end{equation}
we find the coupling of the R-R potential to the (already GSO projected) 
boundary state for an individual strand \eq{eq:RRB} to be
\bea \label{eq:RRcoupling}
\cA_{\rm R, \psi}^{(s)} &=& \braket{\hat{C}_{(n)}}{D5;f_{(s)}} _{\psi,0} \cr
&=& \frac{-i}{4\sqrt{2}\,n!}
\tr \left[ 
\G_{\m_n \cdots \m_1}  \left( \ha \G^{vu}+ \dot{f}_{(s)}^I(v)\G^{Iv} \right) \G^{5678} 
\frac{1+\G_{11} }{2} \right]_{AB} \hat{C}^{\mu_1\ldots\mu_n}~. \qquad  \label{eq:RRcoupling2}
\eea
This then combines 
with the bosonic zero mode part of the amplitude $\cA_0^{(s)}$ given in \eq{eq:A_0^s} 
and we sum over strands to obtain the full R-R amplitude $\cA_{\rm R}$.
We then extract the gauge field profile via 
\be
\hat{C}^{(n)}_{\mu_1\ldots\mu_n}(k) ~=~ \frac{\delta {\cal A}_{\rm R}}{ \delta \hat{C}^{(n) \mu_1\ldots\mu_n}}\quad (\mu_1<\mu_2\ldots<\mu_n)\,,
\ee
and as for the NS-NS calculation we insert the propagator and perform the Fourier transform. The fields which are non-trivial only in the presence of a travelling wave are then
\be \label{eq:RRstring} 
\hat{C}^{(4)}_{vbcd}  
~=~ \frac{Q_5}{\sqrt{2} \k L_T}  
\int\limits_0^{\,\,L_T}
 \!d\hat{v} \,
\frac{  \dot{f}_a \e_{abcd} }{|x_i-f_i(\hat{v})|^2}~,\qquad
\hat{C}^{(6)}_{vi5678}  
~=~ \frac{Q_5}{\sqrt{2} \k L_T} 
\int\limits_0^{\,\,L_T}
 \!d\hat{v}
\,\frac{\dot{f}_i}{|x_i-f_i(\hat{v})|^2} ~
\ee
which agrees with \eq{eq:sugralinear}. This completes the link between the microscopic and macroscopic descriptions of a D5-brane with a travelling wave.

To conclude, we have shown how to derive the supergravity fields sourced by D\nobreakdash-\hspace{0pt}brane/momentum bound states from disk amplitudes, describing in detail the D5-P frame calculation. In the D1-P frame, the computation proceeds exactly as for the $\mR^4$ directions of the D5-P system and the results may be readily obtained by changing the index $i \to I$ and the charge $Q_5 \to Q_1$ where appropriate. We have seen that our calculations reproduce the full form of the harmonic functions in the known supergravity solutions, which was not possible in the analogous calculation in the D1-D5 duality frame~\cite{Giusto:2009qq}. This is due to the fact that in the D$p$-momentum
duality frames the profile function parametrizing the solutions
arises as a condensate of massless open strings related to the
physical shape of the D-brane, which we were able to treat exactly using the boundary state formalism.
We hope that combining this result with the analysis of the D1-D5 bound state
will provide a way to study the three-charge D1-D5-P system using string amplitudes.
Work in this direction is in progress. \\

\vspace{2mm}
\noindent {\large \textbf{Acknowledgements} }

\vspace{2mm}
We thank Stefano Giusto, Vishnu Jejjala, Francisco Morales and Daniel Thompson for fruitful discussions.
WB and DT are supported by STFC studentships.

%

\providecommand{\href}[2]{#2}\begingroup\raggedright\endgroup

\end{document}